\documentclass{elsart}

\def\ee #1 {\times 10^{#1}}          % \ee p       10^p
\def\ut #1 #2 { \, \mathrm{#1}^{#2}} % \ut unit p  unit^p
\def\u #1 { \, \mathrm{#1}}          % \u unit     unit

\def\pdeg           {$.\kern-.25em ^{^\circ}$}
\def\degree{\ifmmode{^\circ} \else{$^\circ$}\fi}
\def\ee #1 {\times 10^{#1}}          % \ee p       10^p
\def\ut #1 #2 { \, \textrm{#1}^{#2}} % \ut unit p  unit^p
\def\u #1 { \, \textrm{#1}}          % \u unit     unit

\usepackage{natbib}

\usepackage{graphicx}
\usepackage{amssymb}

\begin{document}

\begin{frontmatter}



\title{A Radio Study of the Mouse, G359.23--0.82}


\author{F. Yusef-Zadeh}
\address{Dept. Physics and Astronomy, Northwestern University, 2145 Sheridan
Road, Evanston IL. 60208, USA }

\author{B. M. Gaensler}
\address{Harvard-Smithsonian Center for Astrophysics, 60 Garden Street
MS-6, Cambridge, MA 02138, USA }

\begin{abstract}
% Text of abstract

The recent detection of a young pulsar powering ``the Mouse'', 
G359.23--0.82, as well as detailed imaging of surrounding nebular X-ray 
emission, have motivated us to investigate the structural details and 
polarization characteristics of the radio emission from this axisymmetric 
source with a supersonic bow shock. 
Using polarization data at 3.6 and 6cm, we find that the magnetic field 
wraps around the bow shock structure near the apex of the system, but 
downnstream runs parallel to the inferred direction of the pulsar's 
motion. The rotation measure (RM) distribution of the Mouse also suggests 
that the low degree of polarization combined with a high RM ahead of the 
pulsar result from internal plasma within the bowshock region. In 
addition, using sub-arcsecond radio image of the Mouse, we identify 
modulations in the brightness distribution of the Mouse that may be 
associated with the unshocked pulsar wind behind the pulsar. Lastly, we 
discuss the relationship between the Mouse and its neighboring shell-type 
supernova remnant G359.1--0.5 and argue that these two sources could 
potentially have the same origin.

\end{abstract}


\end{frontmatter}

% main text
\section{Background}
\label{}

G359.23--0.82 (``the Mouse''), with its long axisymmetric nonthermal 
nebula extending for 12~arcminutes, was first discovered as part of a Very 
Large Array (VLA) radio continuum survey of the Galactic center at 20cm 
(Yusef-Zadeh and Bally 1987).  
A bow-shock structure was noted  
along the  eastern edge of the  nebula (Yusef-Zadeh and Bally 
1989).
In addition,  radio continuum data show linearly 
polarized emission from the full extent of the nebula and the spectral 
index distribution between 2, 6 and 20cm remains flat at the head of 
the nebula but steepens in the 
direction away from the head of the Mouse (Yusef-Zadeh and Bally 1989).  
The detection of X-ray emission from this source and the identification of 
a young radio pulsar G359.23--0.82 at the head of the nebula resulted in a 
breakthrough in our understanding of what powers this source (Predehl and 
Kulkarni 1995; Sidoli et al. 1999; Camilo et al. 2002). More recently, 
{\em Chandra}\ observations show detailed structural and spectral 
characteristics of this bow-shock pulsar wind nebula (PWN) associated with 
the Mouse (Gaensler et al. 2004). Modeling of the X-ray emission suggests 
a high space velocity $\approx$600 km~s$^{-1}$ in a relatively warm phase 
of the ISM in order to explain the cometary morphology of this source.

The region where the Mouse is found contains a number of other nonthermal 
radio continuum sources. Figure~1 shows a large-scale 20cm view of this 
region where two shell-type supernova remnants (SNRs) G359.1--0.5 and 
G359.1--0.2, and a southern segment of a nonthermal radio filament 
G359.1--0.2, known as the ``Snake'', are distributed (Reich \& F\"urst 
1984; Gray et al. 1991; Yusef-Zadeh, Hewitt and Cotton 2004). The 
shell-like SNR G359.1-0.5 has a diameter of about 20$'$ and the extension 
of the tail of the Mouse appears to be pointed toward the center of this 
remnant.  Here, we present high-resolution radio images showing new 
structural details of the Mouse. Using polarization data, we present the 
distribution of the magnetic field and the rotation measure of the 
bow-shock region of the nebula. We also argue that the Mouse and 
SNR~G359.1--0.5 are potentially  associated with each other.

\section{Observations and Results}

Using the VLA in its BnA array configuration, we reduced the archival data 
taken in 1999~October at 3.6cm wavelength. Standard calibrations were 
carried out using 3C~286 and IERS~B1741--312 as amplitude and phase 
calibrators, respectively.  We also calibrated and combined the 6 and 2cm 
data taken in 1987~November and 1988~February in the BnA and CnB array 
configurations, respectively. 3C~286 and NRAO~530 were selected as the 
phase and amplitude calibrators for the 6 and 2cm data. Using different 
robust weighting to the {\it uv} data, the final images were constructed after 
phase self-calibration was applied to all 6, 3.6cm and 2cm data. The 
spectral index distribution of the Mouse based on these measurements will 
be given elsewhere. We believe that the scale lengths of the features that 
are described here are well-sampled in the ${\it uv}$ plane. In 
particular, the best well-sampled baselines range  between 10 and  550 
k$\lambda$ at 3.6cm and and 5 and 250 k$\lambda$ at  
6cm, respectively.

Figures 2a,b shows the combined polarized and total intensity images of 
the Mouse with a resolution of 2.1$''\times1.9''$ (PA~$=-34^{\circ}$) at 
6cm. The total intensity image of the head of the Mouse shows a cone-shape 
structure within which a bright linear feature with an extent of 
$\sim45''$ 
appears to run along the axis of the symmetry. With the exception of 
depolarizing features, which are displayed as dark patches in Figure 2a, 
similar morphology is seen in both the total and polarized intensity 
images. The overall degree of polarization at 6cm ranges between 10 and 
25\%. Detailed 6cm images of the head of the nebula show that the peak 
polarized and total intensity images are offset from each other suggesting 
that the emission is depolarized  at the outermost portion of the 
bow shock. This 
offset is best shown in Figure 3a,b where a slice is cut along the 
axis of 
the symmetry of the distribution of the polarized and total intensity, 
respectively. A sharp rise in the total intensity of the emission having a 
6--7\% degree of polarization is followed by a peak  emission which 
coincides with a drop 
in the polarized emission; the degree of polarized emission is less than 
1\% at this position. In the region where the total intensity falls 
gradually, the degree of the polarization 
rises to values ranging between 5--10\%. It is clear that either the 
magnetic field is highly 
tangled-up  or that there is a high degree of thermal plasma mixed in with 
nonthermal gas at the head of the nebula where no polarized emission is 
detected.  
A more dramatic picture of this depolarization feature at 3.6cm is shown
in Figure 6a, as discussed   below. 

Figure 4a,b  show grayscale and contour images of 
the head of the Mouse immediately to 
the west of the bright bowshock. 
The bright compact source that is 
suspected to be coincident with the pulsar is clearly resolved in 
these images. 
The sharp  distribution of 
the emission ahead of the pulsar when compared with a more gradual 
distribution of emission behind the pulsar supports for the 
compression of gas and magnetic field in the bowshock region resulting 
from the supersonic motion of the pulsar. 
We note  a 4$'' ``$hole$"$  in the 3.6cm brightness distribution within 
which 
the  emission is three times fainter than the features surrounding it. The 
center of the ``hole$"$ lies near $\alpha, \delta (2000) = 17^h 47^m 
15^s.35, 
-29^0 58' 01''.5$. The non-uniform distribution of emission behind the 
bow-shock can be seen throughout the Mouse at 6 and 3.6cm.  Additional 2 
and 6cm images presented in Figures 5a,b also support the evidence for the 
modulation of the total intensity along the axis of the symmetry. The 
morphology of the emission is complicated but Figures 2 -- 5 show that the 
overall brightness temperature along the axis of the symmetry decreases at 
about 
5$'', 12'', 20''$ and 45$''$ west of the bowshock.

The grayscale distribution of the polarized emission superimposed on 
contours of total intensity at 6 
and 3.6cm are  shown in 
Figures 6a,b, respectively.  The length and the position angle of the line 
segments represent the distributions 
of the polarized emission and the  
polarization vectors rotated by 90 degrees. The distribution of the 
polarization vectors  at these wavelengths,  including our 2cm data 
that are not shown here,  
are similar to each other. 
 The striking feature in these images is the recognition of a depolarizing 
feature separating the distribution of the polarized emission in the 
bowshock region from the region behind the pulsar. The position angle of 
the polarization angle vectors changes by 90$^0$ between these two regions.  
The slice representations of the total and polarized intensity at 6cm, as 
shown in Figure 3a,b, are consistent with a picture that the scale of the 
magnetic field within the beam must have changed in order to produce the 
depolarizing feature.  The distributions of the polarization vectors at 
3.6 and 6cm are  used to determine the rotation measure (RM) distribution 
of the polarized emission ahead of the Mouse.

Figure 7a,b show slice representations of the distribution of the RM and 
its error between 3.6 and 6cm cut along the symmetry axis of the Mouse.  
The RM value is on order of 500$\pm$40 rad m$^{-2}$ near the bow shock 
to --400$\pm$20 rad~m$^{-2}$ at the position of the depolarizing feature 
before it increases to a value 0$\pm$60~rad~m$^{-2}$ away from the bow 
shock.  The low value of the RM and its error in the region behind the 
pulsar suggests the intrinsic magnetic field traces the direction of the 
inferred motion. We believe that the Faraday rotation toward the Mouse due 
to interstellar material along the line of sight is well represented in 
the RM distribution behind the pulsar and along the tail of the Mouse. 
However, the high RM values as well as the evidence in field reversal and 
the low degree of polarization ahead of the pulsar are distinctly 
different than the region behind the pulsar.  These effects, in 
particular, the increase in RM toward the head of the Mouse by two orders 
of magnitude imply that Faraday rotation along the line of sight can not 
explain the unusual characteristics of polarization ahead of the pulsar. 
We suggest a picture in which there is a mixture of thermal and nonthermal 
plasma coexisting in the bowshock region. Although, it is expected that 
the magnetic field gets compressed and becomes more uniform in the 
bowshock region of a pulsar wind nebula, internal thermal plasma can 
produce a high degree of Faraday rotation and depolarization (Burn 1966).  
Future detailed polarization observations using multiple bands in order to 
determine the true distribution of Faraday rotation should be able to test 
this interpretation.

Using the dispersion measure of $\approx$100 cm$^{-3}$ pc (Camilo et al. 2002) and 
an upper  limit of  
RM$\approx$ 100 rad m$^{-2}$, the parallel component of the magnetic field 
along the line of sight is estimated to be $\approx1.2 \mu$G. This unusually low 
value 
of the magnetic field along the line of sight is likely to be 
underestimated since the sign of the RM could change between the head and 
the tail of the Mouse. The fluctuation of electron density and magnetic 
field of a turbulent medium along the line of sight, as well as from the 
region within the nebula, can severely affect the correct estimate of the 
magnetic field along the line of sight (Beck et al. 2003). In spite of 
this difficulty, the RM estimate made here is at least an order of 
magnitude less than the estimate made toward objects lying within a few 
hundred pcs of the Galactic center (e.g., Yusef-Zadeh, Wardle and 
Parastaran 1997).  The RM value estimated from this 
study is consistent with a source lying less than 5 kpc from us.

\section{Discussion}

One of the most puzzling aspects of radio observations reported here is 
the distribution of synchrotron emission near the apex of the pulsar wind 
where the intensity is modulated at 3cm. The drop in 
synchrotron emissivity may be accounted for in terms of the 
unshocked wind arising from the pulsar or an inhomogeneous  
distribution of synchrotron emission.  The numerical simulations carried 
out by Gaensler et al. (2004) have in fact predicted a closed elongated 
structure associated with the termination shock behind a supersonically 
moving pulsar. A more detailed comparison of radio and X-ray data is 
needed to determine the nature of the intensity distribution
and how it may be 
related to the termination shock behind the pulsar.

Radio and X-ray observations suggest the the Mouse could lie at distance ranging 
between 2 and 5 kpc and that this object is not associated with SNR G359.1--0.5.  
Uchida et al. (1992) considered that G359.1--0.5 is associated with a molecular 
ring lying near the Galactic center requiring multiple supernova explosions to 
account for an energy of 6$\times10^{52}$ ergs. However, a subsequent detection of 
maser emission at --5~km~s$^{-1}$ surrounding this large-scale source (Yusef-Zadeh 
et al. 1995), as well as the detection of low value of rotation measure toward the 
remnant (Yusef-Zadeh et al. 2005, in preparation), suggest that SNR G359.1--0.5 is 
unlikely to be located within a few parsec of the Galactic center at the distance 
of 8.5 kpc. One possibility is that the Mouse and SNR G359.1--0.5 are associated 
with each other. The support for this suggestion may come from the absorption 
column measured ranging between $3\times10^{22}$~cm$^{-2}$ and 
$(6\pm2)\times10^{22}$~cm$^{-2}$ toward the remnant (Bamba et al. 2000; Egger and 
Sun 1998) and $(2.7\pm0.1)\times 10^{22}$~cm$^{-2}$ toward the Mouse (Gaensler et 
al. 2004). Depending on how one fits the X-ray spectrum, these estimates could be 
consistent with each other. If SNR G359.1--0.5 and the Mouse lie at the distance of 
5 kpc, the time that it takes for the pulsar to travel 34 pc, the distance between 
the center of the remnant and the pulsar, is $\approx5.7\times10^4$ years. This is 
more than twice the characteristic age of the pulsar $2.5\times10^4$~yr. Recent 
studies suggest that the true age of young pulsars can be longer than their 
characteristic age, so it is possible that the Mouse and SNR G359.1--0.5 could have 
the same origin.

In conclusion, we have presented new high-resolution radio continuum 
images of the Mouse, showing a modulation of the intensity behind the 
pulsar.   
A polarization image of the Mouse at 3.6cm shows that 
 the overall distribution of 
the inferred magnetic field is parallel to the axis of symmetry away 
from the bow shock.  
However, a high degree of depolarization, field reversal and high RM are  
detected ahead of the pulsar near  the interface where the 
polarization angle changes by $90^\circ$. These images indicate that the 
direction of the magnetic field in the bow shock is likely to be tangent 
to the shock normal, though more detailed radio observations are required 
to correct for Faraday effects. We have also argued 
that the Mouse and an adjacent SNR G359.1--0.5, may be associated. Future 
radio and X-ray observations should be able to shed light on the nature of 
the Mouse and SNR G359.1--0.5 as well as their physical relationship.

\section*{Acknowledgments}

The National Radio Astronomy Observatory is a facility of the National
Science Foundation operated under cooperative agreement by Associated
Universities, Inc.  BMG acknowledges the support of NASA through SAO
grant GO2-3075X and LTSA grant NAG5-13032. FYZ was funded 
by NSF AST-03074234 and NASA NAG-9205. We also thank M. Wardle for useful 
discussion.

\vfill\eject

\begin{figure}
\caption{A complete shell-type SNR,
 G359.1--0.5, an incomplete shell-type SNR G359.0--0.9, and the Mouse, 
G359.23--0.82,
are shown at 20cm with a resolution of
12.8$''\times8.4''$ (PA~$=56^\circ$).  
The southern extension of the Snake filament, G359.1--0.2,  is also 
noted (Yusef-Zadeh et al. 2004). 
}
\end{figure}

\begin{figure}
\caption{The top (a) and bottom (b)   panels show grayscale polarized
and total intensity images of the Mouse 
at 6cm, respectively,  based on combined BnA and CnB-array configurations 
with  a resolution of 
2.1$''\times1.9''$ (PA~$=-34^\circ$) and uniform {\it uv} weighting. 
}
\end{figure}

\begin{figure}
\caption{
The top (a) and bottom (b)  panels show a  slice cut  along 
the axis of the symmetry of the respective polarized and total intensity 
images that are  shown in Figure 2. The position angle of 
slice is -90$^{\circ}$.}
\end{figure}

\begin{figure}
\caption{(The top (a) and bottom (b) panels show a
 grayscale total intensity image 
and contours of total intensity 
at (2, 3, 4, 5, 7, 9, 11, 13, 20, 30 and 40) $\times$ 90  $\mu$Jy
beam$^{-1}$ with a
robust weighting of -7 and a
convolved spatial resolution of $0.82''\times0.82''$ 
at 3.6cm, respectively. 
}
\end{figure}

\begin{figure}
\caption{The top (a) and bottom (b)  panels  show  
contours of total intensity 
at 0.5, 1, 2, 3, 5, 7, 10, 15 and 20 mJy beam$^{-1}$ 
and 0.5, 0.75, 1, 1.5, 2, 2.5, 3, 4, 5, 6, 7 and 8 mJy beam$^{-1}$ at 2 
and 6cm, respectively.  The 
spatial resolutions of 2 and 6cm images are  $3''\times1.75''$ 
and 1.1$''\times1.1''$, respectively. The 6cm data are based only on the BnA array 
configuration  data with a robust weighting -7 
whereas the 2cm data have used the data combined from BnA and CnB array 
configuration observations using a  {\it uv} taper 100 k$\lambda$ and a robust 
weighting 4.}
\end{figure}

\begin{figure}
\caption{[Top] The distribution of  the inferred magnetic field 
orientation at 3.6cm with no Faraday correction. 
The length of the vectors correspond to the 
strength of polarized emission. The grayscale image shows the
distribution of  the total intensity with  contours 
set at 0.1, 0.3, 0.5, 0.7, 1, 3 and 5 mJy beam$^{-1}$ and 
 resolution of
$1.4''\times0.7''$ (PA~$=18^\circ$) with a robust weighting 5. [bottom] Similar to 
the top panel except
that this image is based only on BnA array at 6cm 
with a resolution of $1.6''\times1.4''$ (PA~$=-18^\circ$) with uniform 
{\it uv} weighting  and 
contour 
levels 1, 2 and 8  mJy beam$^{-1}$. 
}
\end{figure}

\begin{figure}
\caption{[Top] A slice cut along the symmetry axis of the 
distribution of the rotation measure based on 6 and 3.6cm polarization 
measurements. [Bottom] Similar to the top panel except that the slice 
is made through the error map of the RM distribution. 
}
\end{figure}

\end{document}